\makeatletter
\@ifundefined{@parse@version@dash}{%
\def\@parse@version#1{\@parse@version@0#1}
\def\@parse@version@#1/#2/#3#4#5\@nil{%
\@parse@version@dash#1-#2-#3#4\@nil}
\def\@parse@version@dash#1-#2-#3#4#5\@nil{%
  \if\relax#2\relax\else#1\fi#2#3#4 }
}{}
\makeatother

\documentclass[
aps,
pre,
superscriptaddress, 
singlecolumn,
showpacs,
amsmath,amssymb,
]{revtex4-2}

\usepackage{graphicx}
\usepackage{dcolumn}
\usepackage{bm}

\begin{document}


\title{Optimization of coupling and global collapse in diffusively coupled socio-ecological resource exploitation networks}

\author{Tanja Holstein}
 \affiliation{FutureLab on Game Theory \& Networks of Interacting Agents, Complexity Science, Potsdam Institute for Climate Impact Research, Member of the Leibniz Association, P.O. Box 60 12 03, 14412, Potsdam, Germany}
 \affiliation{Department of Physics, Humboldt University, Newtonstr. 15, 12489 Berlin, Germany}
 \affiliation{Section eScience (S.3), Federal Institute for Materials Research and Testing, Unter den Eichen 87, 12205 Berlin, Germany}
\author{Marc Wiedermann}%
 \email{marcw@physik.hu-berlin.de}
 \affiliation{FutureLab on Game Theory \& Networks of Interacting Agents, Complexity Science, Potsdam Institute for Climate Impact Research, Member of the Leibniz Association, P.O. Box 60 12 03, 14412, Potsdam, Germany}
\author{J\"urgen Kurths}
  \affiliation{Department of Physics, Humboldt University, Newtonstr. 15, 12489 Berlin, Germany}
 \affiliation{Complexity Science, Potsdam Institute for Climate Impact Research, Member of the Leibniz Association, P.O. Box 60 12 03, 14412, Potsdam, Germany}
 \affiliation{Lobachevsky University of Nizhny Novgorod, Nizhnij Novgorod 603950, Russia}


\date{\today}

\begin{abstract}
Single- and multi-layer complex networks have been proven as a powerful tool to study the dynamics within social, 
technological, 
or natural systems.
An often observed common goal there is to optimize these systems for specific purposes by minimizing certain costs 
while maximizing a desired output. 
Acknowledging that especially real-world systems from the coupled socio-ecological realm are highly intertwined
this work exemplifies that in such 
systems the optimization of a certain subsystem, e.g., to increase the resilience against external pressure in an ecological network, may unexpectedly diminish the stability of the whole coupled system. 
For this purpose we utilize an adaptation of a previously proposed conceptual bilayer network model composed of 
an ecological network of diffusively coupled resources co-evolving with a social network of interacting agents that harvest these resources and learn each other's strategies depending on individual success. 
We derive an optimal coupling strength that prevents a collapse in as many resources as possible if one assumes that the agents' strategies remain constant over time. However, we then show that if agents socially learn and adapt strategies according to their neighbors' success, this optimal coupling strength is revealed to be a critical parameter above which the probability for a global collapse in terms of irreversibly depleted resources is high -- an effect that we denote the \textit{tragedy of the optimizer}. 
We thus find that measures which stabilize the dynamics within a certain part of a larger co-evolutionary system may unexpectedly cause the emergence of novel undesired globally stable states. Our results therefore underline the importance of holistic approaches for managing socio-ecological systems because stabilizing effects which focus on single subsystems may be counter-beneficial for the system as a whole. 

\end{abstract}

\maketitle

%

\section{Introduction}
Complex networks have been proven as a powerful framework to study the structure and dynamics in a broad range of real-world systems, ranging from 
social networks~\cite{dodds2003experimental, vosoughi2018spread,stopczynski2014measuring, schleussner2016clustered} to complex adaptive systems in socio-ecology~\cite{preise2018social} and multilayer hierarchical structures in infrastucture~\cite{kivela2014multilayer}, economy~\cite{wiedermann2013node} or even the climate system~\cite{donges2011investigating, feng2012three}. As such they are particularly useful to study meso- and macroscopic emergent phenomena arising from the microscopic interaction between dynamics at individual nodes, such as the stability of power grids~\cite{auer2016impact, nishikawa2015comparative}, survival and co-existence in resource networks or food-webs~\cite{dunne2006network, pilosof2017multilayer, knebel2013coexistence}, 
or the synchronization in networks of coupled oscillators~\cite{arenas2008synchronization, rodrigues2016kuramoto, rakshit2018emergence, maksimenko2016excitation}. 
Specifically, complex networks have also been utilized to analyze a broad range of social dynamics~\cite{castellano2009statistical} and spreading processes~\cite{de2016physics} that fostered the development of associated conceptual models with foci on complex contagion~\cite{dodds_generalized_2005,guilbeault_complex_2018}, opinion dynamics~\cite{watts_influentials_2007, klamser_zealotry_2016, holme_nonequilibrium_2006}
or epidemic spreading~\cite{castellano2010thresholds, pastor2015epidemic, gross2006epidemic}.

Through a combination of techniques from social and ecological network models~\cite{wiedermann2015macroscopic, barfuss_sustainable_2017, geier2019physics, auer2016impact} complex networks have recently been proven as a promising approach to bridge theoretical physics and efforts to understand future trajectories of the Earth system in the Anthropocene~\cite{filatova2013spatial, donges2020earth, schlueter2012new}, where human social dynamics have become a dominant geological process~\cite{steffen2018trajectories, crutzen2006anthropocene}. These so-called {\em World-Earth models}~\cite{strnad2019deep, donges2018taxonomies, donges2020earth} have for example been used to study emergent characteristics of interactions between social networks of resource harvesting agents~\cite{barfuss_sustainable_2017, geier2019physics, wiedermann2015macroscopic}, impacts of multi-agent social learning and market dynamics on deforestation rates in rain forests~\cite{muller2019can}, or the emergence of sudden regime shifts in socio-ecological systems driven by specific network characteristics~\cite{sugiarto2015socioecological, sugiarto2017emergence}. 

One commonly observed goal in managing such socio-ecological systems is their optimization for a specific purpose by minimizing certain costs and maximizing a desired output~\cite{schluter2017framework, young2006globalization} even though such strategies may not necessarily yield desired long-term stable states~\cite{barfuss2018optimization}. Such optimization schemes include for example quota managements in cross-national fisheries~\cite{karagiannakos1996total} control programs to slow down or prevent the spread of harmful species~\cite{carrasco2012towards} or the most cost-effective choice of biodiversity conservation hotspots~\cite{myers2000biodiversity}. However, such strategies often focus primarily on understanding the underlying ecological system and its natural dynamics for establishing an optimal management while the dynamics in the social system and human behaviour that interacts with these systems is neglected or assumed static, thus, posing a key uncertainty in assessing the effectiveness of such measures~\cite{fulton2011human, schluter2017framework}.

Here we exemplify on this potential drawback and show that in a coupled socio-ecological system the optimization of the natural component alone may unexpectedly diminish the stability of the system as a whole and even may cause the existence of undesired stable fixed points corresponding to a global collapse. For this purpose we utilize a recently proposed conceptual socio-ecological bi-layer network model consisting of a social layer and an ecological layer~\cite{wiedermann2015macroscopic, barfuss_sustainable_2017, geier2019physics}. The ecological layer is comprised of a set of nodes that represent an abstract stock of a renewable resource~\cite{perman2003natural}. Each node in the social layer interacts with exactly one of the stocks through different exploitation strategies. These strategies either yield high short-term gain at the cost of depletion or lower short-term gain for the sake of sustained long-term harvest as the stock approaches a positive stable fixed point~\cite{wiedermann2015macroscopic}. Nodes within the social network interact with each other by learning different strategies depending on the differences in their immediate harvesting payoffs. In a substantial addition to previous implementations of this model, we specifically acknowledge that resources do not evolve in isolation, but are often interacting via diffusive coupling, such as in migratory patterns of different mammals~\cite{okubo2013diffusion}, insects~\cite{dobzhansky1943genetics, tufto2012estimating} or fish~\cite{crittenden1994diffusion, radinger2017assessing}. Hence, we treat stocks in the ecological system as connected in a complex network and investigate the effect of diffusion strengths between nodes on the overall stability of the system, as such processes can give rise to rich dynamics ranging from increased synchronizability~\cite{motter2005network} to the emergence of chaos~\cite{nakao2009diffusion}.

Along the aforementioned lines of optimal resource management, we first derive an optimal coupling strength such that a minimum number of stocks reaches an undesired stable fixed point of almost or full depletion. We therefore assume that no interaction and learning in the social layer takes place, thus emulating the neglection of human behaviour in determining an optimal management strategy~\cite{fulton2011human, schluter2017framework}. We then show that under myopic social learning dynamics, where nodes aim to optimize their short term yield, this optimal coupling strength corresponds to a critical value above which a collapse of the entire socio-ecological system becomes likely, an effect that we denote here as {\em the tragedy of the optimizer}. Our findings imply that approaches for managing socio-ecological systems that focus on enhancing the stability in single subsystem may at the same time be counter-beneficial for the system as a whole. 
This calls for a further investigation of human behaviour and decision making in shaping trajectories and stability of coupled socio-ecological systems and, thus, conceptually underlines the importance of holistic approaches for determining appropriate management strategies.

The remainder of this paper is organized as follows. Section~\ref{sec:model_descript} presents the specifics of the model that is used in this work. We then present the corresponding results in Sec.~\ref{sec:results}. Specifically, we first derive in Sec.~\ref{resource_eq} an optimal coupling strength for the ecological network if no social interactions take place. Sec.~\ref{sec:with_social_learning} studies the stability of our coupled socio-ecological model if there are social interactions and assesses the influence of interaction rates and diffusive coupling strength on the existence of desired and undesired equilibria. Ultimately, Section~\ref{sec:optimal_coupling_is_critical} demonstrates that the derived optimal coupling strength directly corresponds to a critical coupling strength above which the entire system is likely to collapse. We conclude our work in Sec.~\ref{sec:conclusion} with a summary of the results and an outlook to future work.

\section{Model description}
\label{sec:model_descript}
We model the interaction of a stylized social system with diffusely coupled individual resources by means of a bi-layer network $G$. $G$ consists of a \textit{resource} layer $G^R(V^R,L^R)$ with $N$ nodes ${V^R=\{ v_1^R,v_2^R,...,v_N^R\}}$ and $M^R=|L^R|$ edges and a \textit{social} layer $G^S(V^S, L^S)$ with $N$ nodes $V^S=\{ v_1^S,v_2^S,...,v_N^S\}$ and $M^S=|L^S|$ edges. Nodes $v_i^S$ represent agents $i$ that interact with their own renewable (abstract) resource stocks $s_i$ at the corresponding nodes $v_i^R$.

\subsection{Node dynamics and diffusion}

We first describe the node dynamics in the resource layer $G^R$. Here, each node $v_i^R$ represents a time-dependent renewable stock $s_i(t)$ whose dynamics are modeled by a logistic growth function with linear extraction, \cite{fisheries}:
\begin{equation}
    \frac{ds_i(t)}{dt}=a_is_i(t)(1-s_i(t)/\kappa_i)-s_i(t)E_i.\label{loggrowth}
\end{equation}
Here, $a_i$ is the growth rate, $\kappa_i$ the capacity and $E_i$ the extraction or harvesting strategy. Equation \ref{loggrowth} is commonly used to describe various real-world resource systems~\cite{perman2003natural} and has  been applied previously in the context of investigating an adaptive social network co-evolving with dynamic node states \cite{wiedermann2015macroscopic}, a multi-layer network setup describing a resource-user system with a governance layer \cite{geier2019physics} and the effects of heterogeneous resource distributions  \cite{barfuss_sustainable_2017}. Specifically, we assume the harvest of each node, given in the last term of Equation \ref{loggrowth} by $h_i = s_iE_i$, to be linearly dependent on the available stock, a common assumption in the context of resource economics~\cite{perman2003natural} or socio-ecological systems modeling~\cite{lade2013regime, cooke_one-dimensional_1986}. In contrast to the assumption of a constant harvest, this functional form accounts for the requirement that if $s_i(t)=0$ the corresponding harvest vanishes as well, $h_i(t)=E_is_i(t)=0$. It has additionally been shown that such linear extraction is the optimal harvesting strategy for resources with logistic growth~\cite{brown1995ecological} and explicitly incorporates that success in harvesting is typically positively correlated with the amount of available resource \cite{fryxell2010resource}. 
We set the growth rate $a_i=a=1$ and the capacity $\kappa_i=\kappa=1$ for all stocks $i=1,\ldots,N$ so that time and stocks are measured in dimensionless quantities. An individual effort $E_i\in \{ E^+, E^-\}$ is assigned to each node and, for the chosen parameters $a$ and $\kappa$, leads to two possible equilibrium states. For a \textit{high} effort $E^{+}>1$, all stocks converge to an empty state $s_0^{+}=0$, implying depletion of the resource, whereas for a \textit{low} effort $0<E^{-}<1$ all stocks converge to $s_0^{-}=1-E^{-}$, preserving the resource while initially providing less harvest. We choose to set the high effort to $E^{+}=1.5$ and the low effort to $E^{-}=0.5$ so that $E^{-}$ maximizes the sustainable equilibrium harvest $h_0^{-}=E^{-}\cdot(1-E^{-})$ and both efforts are distributed symmetrically around $E=1$.  
 
As most entities in real-world ecological systems do not evolve in isolation, the individual renewable stocks $s_i(t)$ are coupled through Laplacian diffusion along the links $L^R$ of the network. Hence, by using the Laplacian operator $\hat{L}$ \cite{fiedler1989laplacian} and diffusion rates $\alpha$, the dynamics given by Eq.~\eqref{loggrowth} change to:
\begin{equation}
\label{diffusiveloggrowth}
    \frac{ds_i(t)}{dt}= s_i(t)(1-s_i(t))-E_is_i(t)-\alpha \sum_{j=1}^N L_{ij}s_j
\end{equation}
Here, $\alpha$ can be understood as the rate at which the stock $s_i$ is transferred into the neighboring stocks $s_j$. $L_{ij}$ are the elements of the Laplacian $\hat{L}$, given by $\hat{L}=\hat{D}-\hat{A}$, where $\hat{D}$ is the degree matrix and $\hat{A}$ is the adjacency matrix of stock network graph. 

The introduction of diffusive coupling along the network structure alters the previously mentioned equilibrium stable states of $s_i$ (see above), an effect that we investigate in detail below in Sec.~\ref{resource_eq}.

\subsection{Social learning of exploitation strategies}
\label{sec:social_learning_of_exploitation}
We now describe the dynamics within the social network layer $G^S(V^S, L^S)$ where each node $v^S_i$ represents an agent $i$ exploiting its individual resource stock on the corresponding node $v^R_i$ in the resource layer $G^R(V^R, L^R)$ (see above). The edges $L^S$ represent social ties between the nodes/agents, such as friendships or business partnerships.

In real-world social systems and economies, exploitation strategies rarely remain constant, but may be subject to either rational economic optimization \cite{fisheries} or, for instance, social learning depending on one's own and other's success ~\cite{wiedermann2015macroscopic,barfuss_sustainable_2017,traulsen}. Here, we follow upon previous studies in modeling socio-ecological networks ~\cite{wiedermann2015macroscopic,barfuss_sustainable_2017,geier2019physics} and describe the temporal evolution of individual exploitation efforts $E_i$ through individual activation and social learning.

At first, each agent $i$ becomes active after an individual waiting time $\tau_i$, drawn independently, with its probability density corresponding to an exponential distribution given by
\begin{equation}
\label{poisson}
    p(\tau_i)=\Delta T^{-1}\exp(-\tau_i/\Delta T).
\end{equation}
This typical choice of waiting times in social systems~\cite{haight1967handbook} leads the average waiting time between two social updates of the same agent to be $\langle \tau_i \rangle _i=\Delta T$.

From there, the dynamics in the social system are calculated as follows:
\begin{enumerate}
    \item The system is integrated forward in time according to Eq. \eqref{diffusiveloggrowth} until the minimum of all waiting times $\tau_i$ is reached.
    \item The corresponding agent $i$ with the effort $E_i$ randomly selects an agent $j$ with the effort $E_j$ from its neighborhood and compares the respective efforts.
    \item If $E_i\neq E_j$ the harvest difference ${\Delta h_{ij}=h_j-h_i}$ is computed, corresponding to a comparison between the immediate success of the two employed exploitation efforts $E_i$ and $E_j$. Agent $i$ then adopts the exploitation effort of agent $j$ with a probability given by the logistic function ${p(E_i \rightarrow E_j)=0.5\tanh(\Delta h_{ij}+1)}$, a functional form that has been derived from experimental results \cite{traulsen} and has been successfully employed in previous modelling implementations ~\cite{barfuss_sustainable_2017,wiedermann2015macroscopic} that are similar to the one studied here.
    \item A new waiting time $\tau_i$ is drawn for agent $i$ and the model returns to step 1 above.
\end{enumerate}
Once all agents have chosen either one of the two possible exploitation efforts, the model reaches a steady state at a time $t_f$, since no further changes to individual efforts are possible, ending the iterations.

Note that, in contrast to aforementioned preceding works~\cite{wiedermann2015macroscopic, geier2019physics}, the social network structure studied here is static and does not include adaptive rewiring. As identified previously \cite{wiedermann2015macroscopic}, the social update time $\Delta T$ is an important parameter for the outcome of the system because it sets the relative time scale between the dynamics of the stocks $s_i(t)$ and the social dynamics. In implementations without diffusive coupling the model converges to a state where all agents chose the high effort $E^{+}$, at shorter social updates times, leading to a depletion of all stocks. In contrast, for longer social update times the model converges to a state where all agents chose the low effort $E^{-}$, with the stocks converging to $s_0^{-}=1-E^{-}=0.5$. In fact, highly harvested stocks are depleted by their exploiting agents behavior before their next social update. Thus the agents are not able to anticipate the consequences of their effort choices. 

\begin{figure*}[t]
\includegraphics[width=0.7\textwidth]{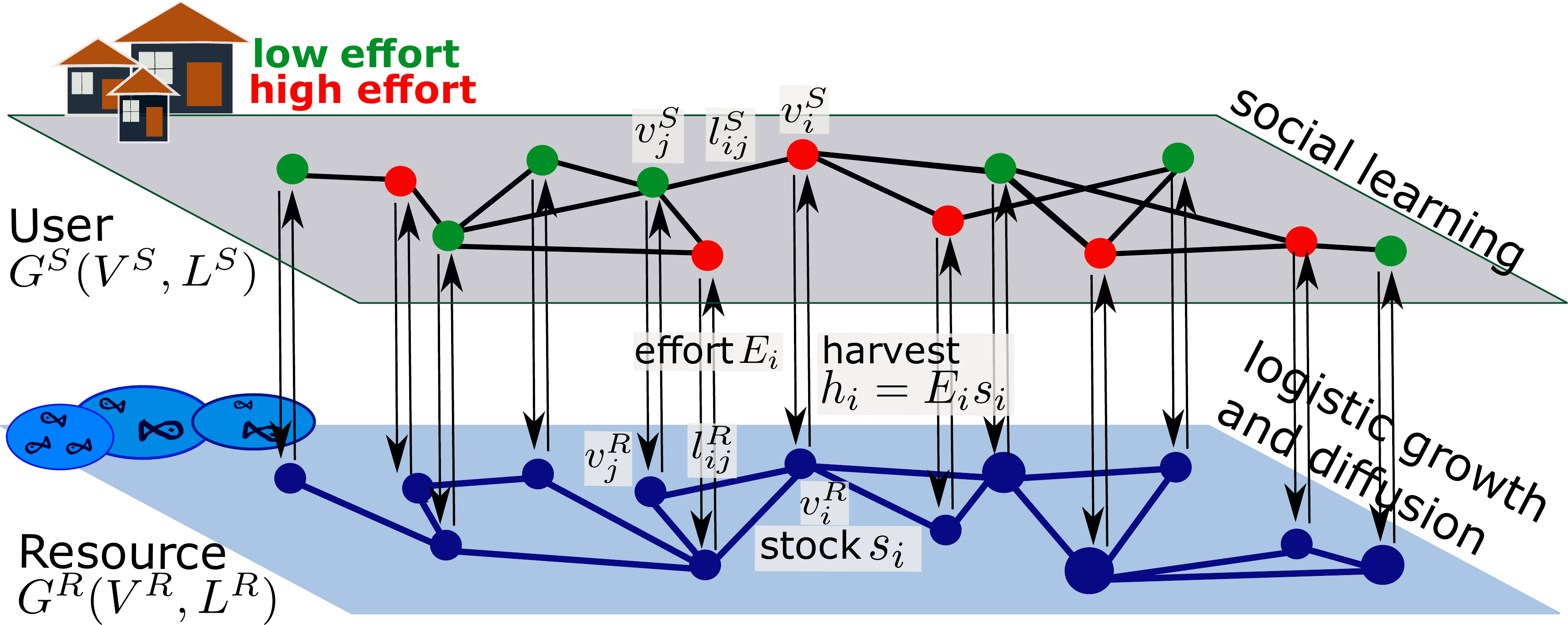}
\caption{\label{model}Schematic visualization of the bi-layer network model that is studied in this work. The upper layer (grey shaded area) represents the social network $G^S(V^S, L^S)$, whose nodes $v^S$ are interpreted as agents, such as individuals or communities, and links $l^S$  indicate social relationships such as friendships or business partnerships between them. Different node colours indicate the two possible effort choices, i.e., high $E^+$ and low $E^-$ effort. Within this social layer, imitation of exploitation behavior takes place through social learning. The bottom layer is the stock network $G^R(V ^R, L^R)$, where individual nodes $v^R$ represent a resource stock. The links $l^S$ represent diffusion pathways between the stocks. Both layers are connected since agents in the social network exploit their respective stocks with efforts $E_i$ that are chosen according to their individual success in obtaining harvest $h_i$.}
\end{figure*}

A schematic representation and summary of the model dynamics and related variables is shown in Fig.~\ref{model}. One network layer corresponds to the social network $G^S$ and the other to the stock network $G^R$. Both networks have an identical set of nodes as each agent in the social network harvests a single stock in the stock network while the edges differ. Over the course of this work, both networks are generated as Watts-Strogatz random networks~\cite{ws_smallworld} with $N=400$ nodes which is consistent with previous works~\cite{barfuss_sustainable_2017, geier2019physics, wiedermann2015macroscopic}. It has been checked that different network sizes do not qualitatively change the results of the model as long as $N$ is of appropriate size, i.e., $N\gtrapprox100$ (not shown). This observation also aligns with previous results from an analytical mean-field approximation of a similar model setup which showed that the model dynamics are expected to be independent of $N$~\cite{wiedermann2015macroscopic}. For the stock network layer, we test the influence of both, average degree $K$ and the rewiring probability $p$, on the expected outcome of the model. The social network layer is consistently set up with an average degree of $K=20$ and a rewiring probability of $p=1$. All stocks start out at full capacity, so that $s_i(t=0)=1$.

\section{Results and discussion}
\label{sec:results}
We now investigate numerical simulations of the network dynamics described in the previous section. We therefore first study the case of an infinite social update time, meaning that no interactions take place between nodes in the social network regardless of their state and harvest. This allows us to gain an understanding of how the diffusive coupling of stocks in the resource layer affects its overall stability and fixed points. From there, we lower the  social updates time to finite numbers in a range that has been studied in previous works~\cite{wiedermann2015macroscopic, barfuss_sustainable_2017, geier2019physics}. We ultimately show that the choice of a coupling strength that is optimal for the resource network without social updates leads to a critical transition in the whole coupled socio-ecological system once social interactions between agents in the social layer are explicitly accounted for.

\begin{figure*}[t]
\includegraphics[width=0.7\textwidth]{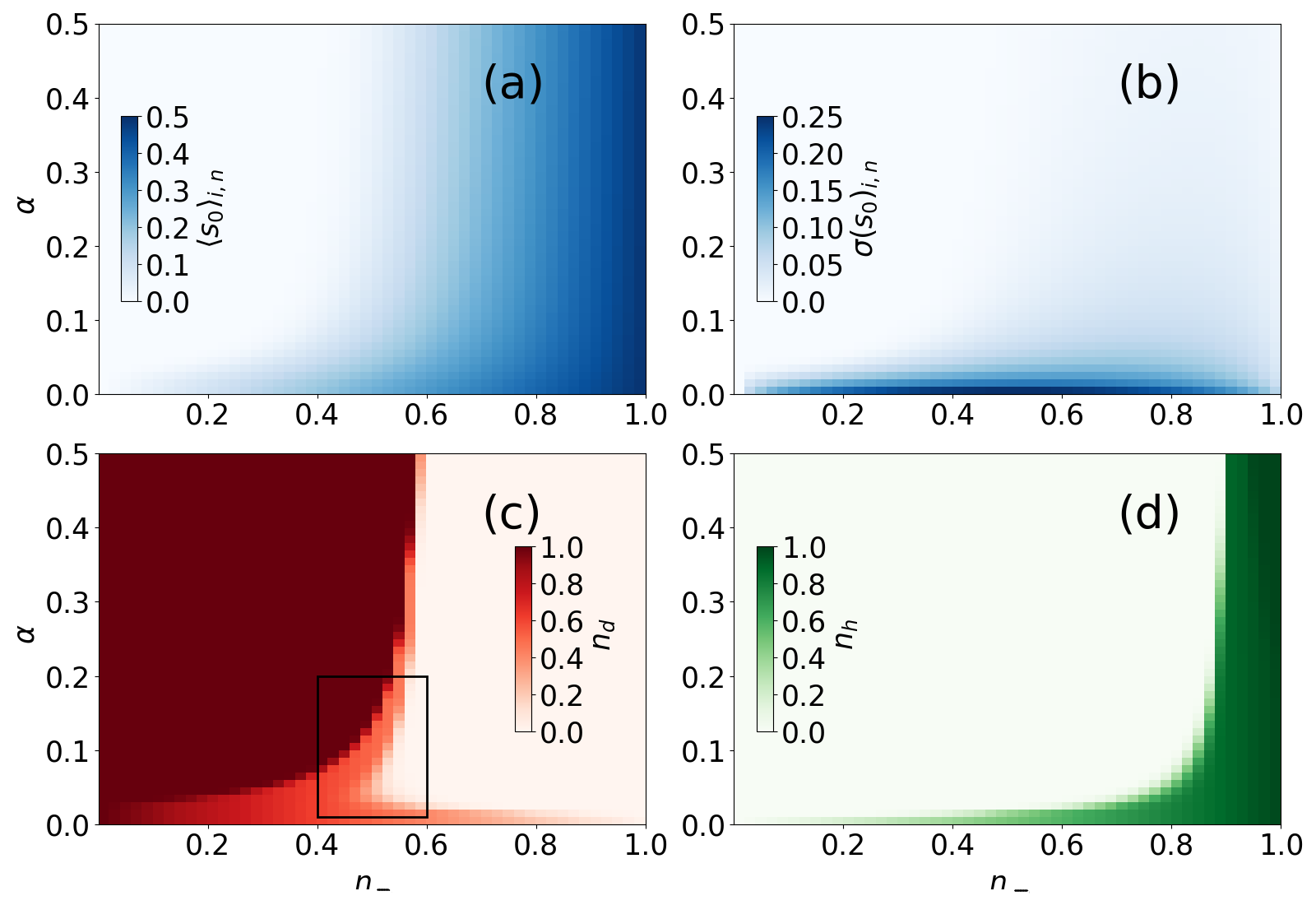}
\caption{\label{fig1}(a) Mean $\langle s_{0} \rangle_{i,n}$ and (b) standard deviation $\sigma(s_0)_{i,n}$ of the equilibrium resource stocks computed over an ensemble of n=200 simulations for different choices of $\alpha$ and $n_-$. (c) Fraction of depleted resource stocks $n_d$, i.e. the fraction of all stocks $s_i$ whose equilibrium value $s_{i,0}$ is lower than 0.1. The black box indicates the area where diffusive coupling appears to reduce the number of depleted stocks compared to the case where no diffusion ($\alpha=0$) takes place. (d) The corresponding fraction of healthy resource stocks $n_h$, i.e. the fraction of all stocks $s_i$ whose equilibrium values $s_{i,0}$ exceed 0.4.}
\end{figure*}

\subsection{Optimization of resource resilience with infinite social update time}
\label{resource_eq}
Our first goal is to understand the effects of diffusive coupling on the stock network's equilibrium states for a fixed number $n_-$ of stocks which are exploited with a low effort. We therefore set the social update $\Delta T=\infty$, so that no social updates occur.

Numerical simulations are performed for different combinations of the coupling strength $\alpha$ and the fraction of nodes $n_-$ exploited with low effort. The $n_-$ nodes that hold the effort $E^-$ are therefore randomly selected at each initialization of the simulation. Figure \ref{fig1}a shows the mean equilibrium stock $\langle s_0 \rangle _{i,n}$ averaged over an ensemble of $n=200$ simulations per set of parameters $\alpha$ and $n_{-}$ as well as all $N=400$ nodes. An increase in the coupling strength until $\alpha=0.2$ leads to a lower mean equilibrium stock, Fig.~\ref{fig1}a. From there, a further increase in the $\alpha$ seems to have only minor effects, Fig.\ref{fig1}a. What we observe is the transition between two separate regimes: In the first regime, for $\alpha\ll0.2$, the evolution of resource stocks is mostly dominated by its intrinsic growth and harvest dynamics. With increasing coupling interactions between resources become stronger until for $\alpha\gg0.2$ the resources are sufficiently mixed. They then exhibit dynamics similar to what one would expect from a single global stock simultaneously available to a all nodes. As such, the second limiting regime can be interpreted as a common pool resource which is a popular model of shared resources in socio-ecological systems\cite{Ostrom419, TAVONI2012152}. The exact value of $\alpha$ (here $\alpha\approx 0.2$) at which this transition occurs (see Fig.\ref{fig1}a) most likely depends on the various parameters of our model, such as the average degree $K$, the levels of exploitation efforts $E_i$ as well as the underlying network topology. Since for the purpose of this work we mainly focus on the qualitative differences between the two observed regimes, a further in-depth quantitative analysis of the transition point remains as a subject of future research. 

We also observe that the introduction of diffusive coupling substantially reduces the standard deviation of the equilibrium stock (Fig.~\ref{fig1}b) for increasing $\alpha$. Hence, we conclude that diffusively coupling the stock network generally leads to a lower mean resource stock combined with a more homogeneous equilibrium stock distribution, facilitated by the flow of stock from nodes being exploited with low effort $E^-$ to nodes being exploited with high effort $E^+$.

In other words, diffusively coupling the renewable stocks thus seems to be detrimental for the network as a whole. However, as described in Section \ref{model}, the stocks exploited with a high effort $E^+$ approach an empty state $s_0^+=0$ in an uncoupled network. They take a very long time to recover from that state even if the exploiting agent changes their exploitation strategy back to the low level $E^-$. Through diffusive coupling stocks from neighbouring nodes exploited with low effort levels $E^-$ flow to highly exploited nodes (with effort level $E^+$) potentially saving them from depletion. In order to investigate such an effect, we define two categories of stocks according to their equilibrium value compared to the sustainable equilibrium stock value of $s_0^-=0.5$. A stock is defined as \textit{irreversibly depleted} if its equilibrium value $s_{0,i}$ is lower than $20\%$ of the sustainable equilibrium stock, meaning $s_{0,i}\leq0.2\cdot 0.5=0.1$. Analogously, an equilibrium stock is defined as being in a \textit{healthy} state if $s_{0,i}$ is higher than $80\%$ of the sustainable equilibrium stock, meaning $s_{0,i}\geq0.8\cdot0.5=0.4$. Additionally studying different percentages from $10\%$--$30\%$ ($70\%$--$90\%$) for defining \textit{depleted} (\textit{healthy}) equilibrium stocks yielded no significant changes in the results for a range of $\pm10\%$ (not shown). 

\begin{figure*}[t]
\includegraphics[width=0.7\linewidth]{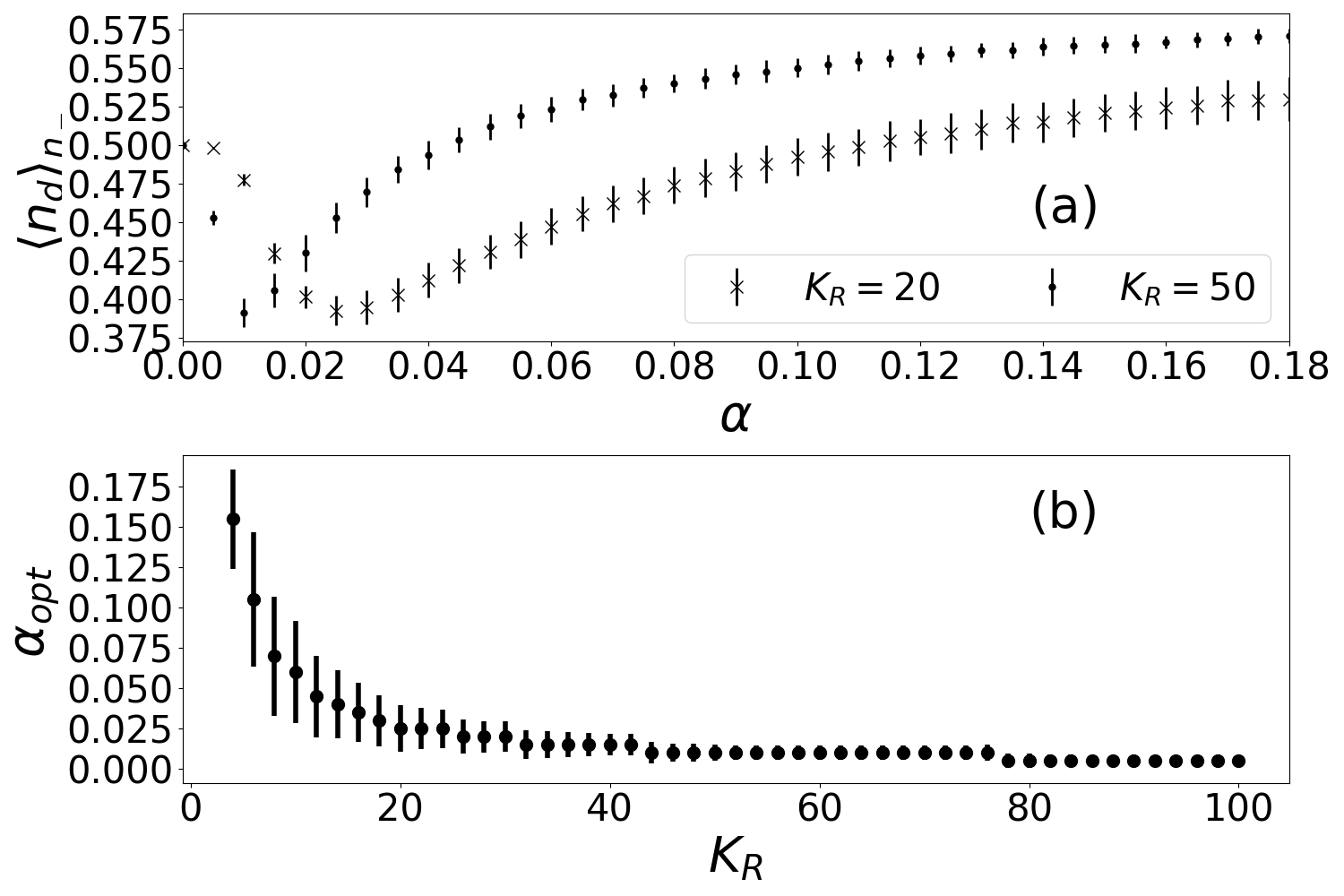}
\caption{\label{fig2} (a) Fraction of depleted nodes $n_d$ as a function of the coupling coefficient $\alpha$ for a resource degree of $K_R=50$ (dots) and $K_R=20$ (crosses). The distinct minimum for $\langle n_d \rangle $ indicates the optimal coupling coefficient $\alpha_{opt}$. (b)  $\alpha_{opt}$ as a function of $K_R$. The error indicates one half of the full-width-half-minimum (FWHM) of the corresponding curve.}
\end{figure*}

Following the above definition, the fraction of depleted ($n_d$) and healthy stocks ($n_h$)  is computed for varying choices of $\alpha$ and $n_-$,  We find that $n_h$ continuously decreases with increasing $\alpha$, since any form of coupling leads to the outflow of stock towards nodes with lower stock levels, Fig \ref{fig1}d. Generally, we uncover that coupling slightly increases $n_d$ with increasing $\alpha$, Fig.~\ref{fig1}c. However for parameter choices in the region $0.4<n_-<0.6$ and $0.01<\alpha<0.2$ (black box in Figure \ref{fig1}c), we find a decrease in the number of depleted stocks which indicates that there seems to exist an \textit{optimal coupling coefficient} $\alpha_{opt}$ that minimizes the fraction of depleted stocks $n_d$, while only slightly decreasing the fraction of healthy stocks $n_h$. 

In order to estimate $\alpha_{opt}$, we now compute the average fraction of depleted stocks $\langle n_d(\alpha) \rangle _{n_-}$ over all fractions $n_-$ of nodes exploited with low effort.  We find that $\langle n_d(\alpha) \rangle _{n_-}$ shows a distinct minimum (Fig.~\ref{fig2}a) and we define the optimal coupling coefficient $\alpha_{opt}$ to be the value of $\alpha$ where $\langle n_d(\alpha) \rangle _{n_-}$ is minimal. Computing $\alpha_{opt}$ for different choices of rewiring probability $p$ and average degree $K_R$ reveals that only $K_R$ substantially affects the numerical value of $\alpha_{opt}$ (results for $p$ are therefore not shown). The specific numerically obtained dependence between $K_R$ and $\alpha_{opt}$ shows that $\alpha_{opt}$ monotonically decreases with increasing $K_R$, Fig.\ref{fig2}b. 

Thus, for the coupled stock network, there seems to be an \textit{optimal} flow $\alpha_{opt}$ preserving the maximum possible number of stocks from depletion and $\alpha_{opt}$ appears to be inversely proportional to the average network degree $K_R$. As the diffusive flow between the stocks determined by $K_R$ (giving the number of available diffusion pathways) and $\alpha$ (giving the {\em width} of these pathways), there seems to be a universal optimal diffusive flow characteristic for the system given by a properly rescaled product of $K_R$ and $\alpha_{opt}$. This relationship should be further investigated, possibly including varying network topologies.
As will be discussed in the next section, the here obtained $\alpha_{opt}$ is a critical parameter in understanding the behaviour of the coupled socio-ecological system.

\subsection{Effect of the diffusive flow with social learning}
\label{sec:with_social_learning}

\begin{figure*}[t]
\includegraphics[width=0.6\textwidth]{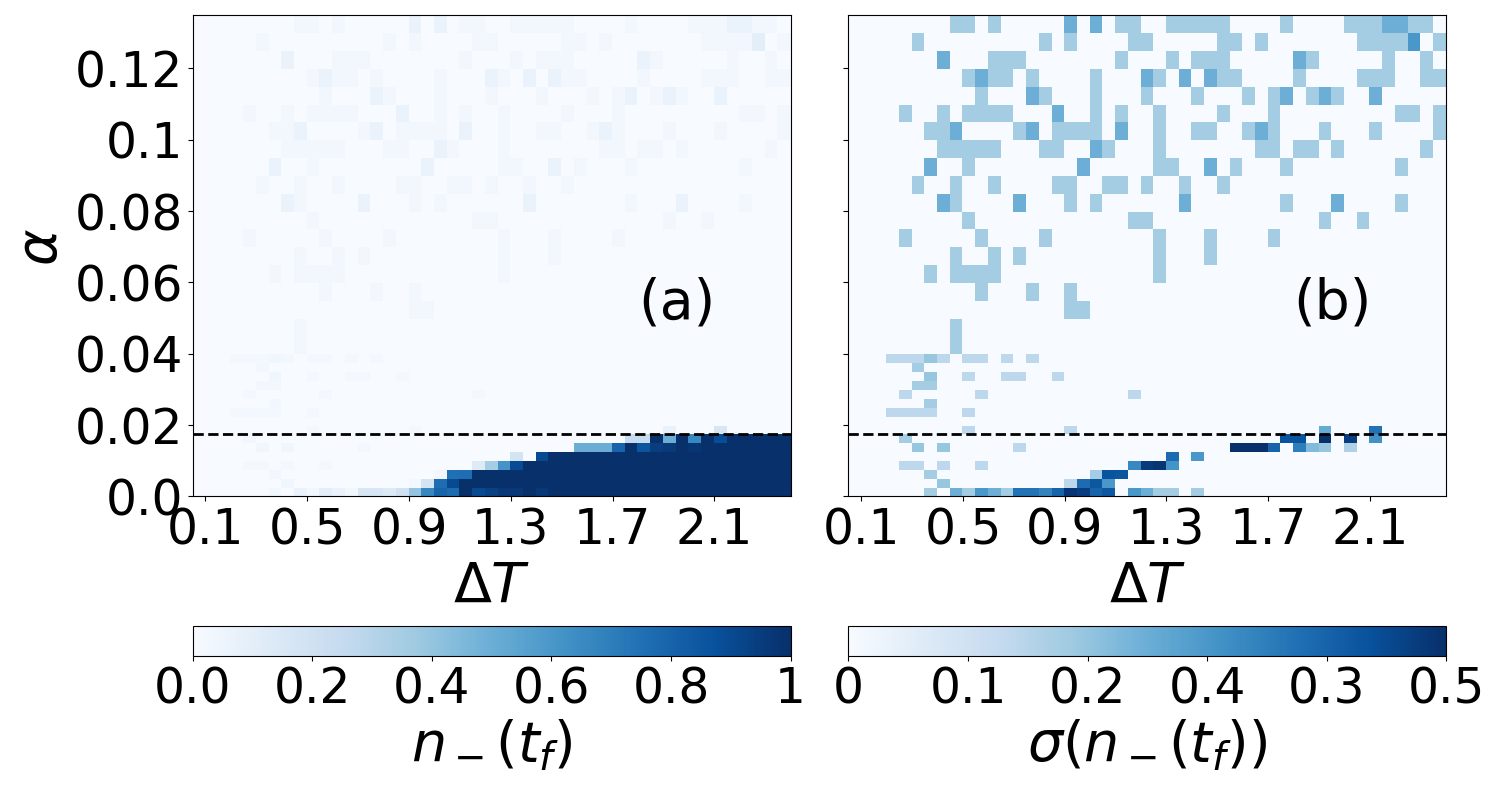}
\caption{\label{fig3}(a) Fraction of agents with low effort at consensus $n_-(t_f)$. Darker shades correspond to more agents choosing a low effort $n_-$. (b) The corresponding standard deviation $\sigma(n_-(t_f))$ at consensus. Again, darker shades correspond to a higher standard deviation. The results are computed from an ensemble of $n=50$ simulations for different choices of $\Delta T$ and $\alpha$. The dashed line marks numerically estimated values of $\alpha=\alpha_{crit}$, above which the agents likely choose a high exploitation effort $E^+$.}
\end{figure*}

We now describe the behavior of the coupled socio-ecological network for varying social update times $\Delta T$ and coupling strengths $\alpha$. The fraction $n_-$ of nodes with low effort is initially set as $n_-(t=0)=0.5$ and changes over time, since agents adapt their effort level through social learning. The $n_-(t=0)=0.5$ nodes in the social network that hold the low (and consequently also the high) effort are randomly selected at the beginning of each simulation. The system is integrated forward until consensus is reached at a time $t_f$, i.e. all agents employ one of the two efforts and therefore no further social updates take place. For each set of parameters we evaluate ensembles of $n=50$ numerical simulations, while we keep the network parameters $p=1$ and $K=20$ fixed for both the social and ecological network. 
\begin{figure*}[t!]
\includegraphics[width=0.9\textwidth]{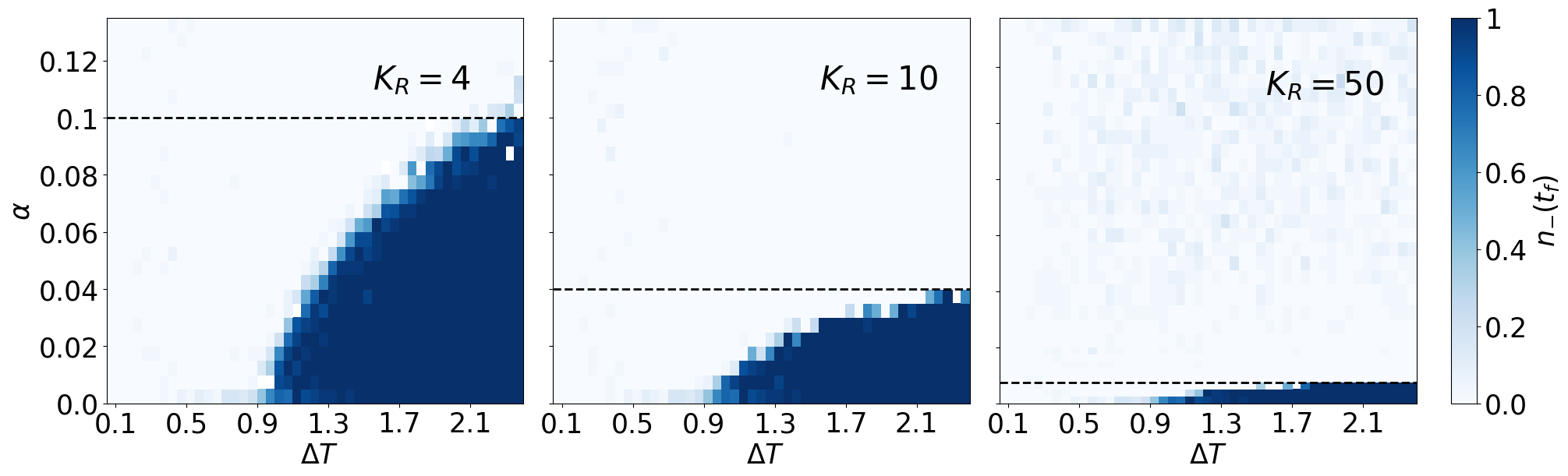}
\caption{\label{fig4}Fraction of agents with low effort at consensus $n_-(t_f)$  computed from an ensemble of $n=50$ network simulations for different choices of $\Delta T$ and $\alpha$. The average degree $K_R$ used for each computation is specified in each panel. Darker shades correspond to more agents choosing a low effort $n_-$. The dashed line marks numerically estimated values of $\alpha=\alpha_{crit}$, above which the agents likely choose an unsustainable exploitation effort $E^+$.}
\end{figure*}
Without diffusive coupling ($\alpha=0$), a transition from all agents choosing a high effort ${\langle n_-(t_f)\rangle=0}$ to all agents choosing a low effort ${\langle n_-(t_f)\rangle =1}$ appears around a critical update time $\Delta T_{crit} \approx 1$, Fig.\ref{fig3}a , consistent with results from previous works \cite{wiedermann2015macroscopic}. With increasing  $\alpha$ the critical update time $\Delta T_{crit}$ increases until at a numerically estimated critical value of $\alpha_{crit}\approx 0.02$ all agents choose a high effort independently of the social update time, Fig.~\ref{fig3}a. Note that $\alpha_{crit}$ is close to the previously described optimal coupling strength $\alpha_{opt}$, compare Fig.~\ref{fig2}a and Fig.~\ref{fig3}a. An intuitive explanation for the existence of this critical coupling strength $\alpha_{crit}$ is the following: Stronger diffusive coupling leads to a more homogeneous stock distribution (see again Fig. \ref{fig1}a and b), effectively {\em protecting} highly exploited stocks from complete depletion. As a consequence, all nodes carry similar amounts of stock whether exploited with low or high effort. Because the harvest $h_i(t)=E_is_i(t)$ depends on both, current effort level and the stock, this leads to a higher harvest for agents employing the high effort $E^+$ when compared to the harvest of agents using the low effort $E^-$. This  counteracts the effect of larger social update times which usually give advantage to agents with low effort as their stocks would in the long run approach larger values than those exploited with high effort if no coupling were to exist. This effect can already be observed at lower $\alpha<\alpha_{crit}$ through the increase of the critical social update time $\Delta T_{crit}$ above which the system approaches a long-term sustainable state, since all nodes employ the low effort (lower right corner in Fig.~\ref{fig3}a). However, for $\alpha>\alpha_{crit}$ the corresponding  $\Delta T_{crit}$ vanishes (or moves to values outside the considered parameter regime, i.e. $\Delta T_{crit}>2.5$) and, hence, causing the high effort to be established along the entire network for all choices of $\Delta T$, Fig.~\ref{fig3}a. This nonlinear behavior related to the optimal coupling strength $\alpha_{opt}$ will be further investigated below in Sec.~\ref{sec:optimal_coupling_is_critical}.

For $\alpha <\alpha_{crit}$, an increase in the corresponding standard deviation $\sigma(n_-(t_f))$ is observed around $\Delta T _{crit}$, Fig.\ref{fig3}b, as is to be expected for systems close to a phase transition~\cite{hohenberg}. Interestingly, for higher values of $\alpha>0.06$ the standard deviation increases again, whereas the agents continue to preferentially choose a high effort, Fig. \ref{fig3}b. This effect can be explained from the fact that the probability for an agent to choose either effort depends on the difference $\Delta h$ between its own and neighbouring agents harvest, ${p(E_i \rightarrow E_j)=0.5\tanh(\Delta h_{ij}+1)}$ (see Sec.~\ref{sec:social_learning_of_exploitation}).  For larger $\alpha$, the highly exploited stocks tend to increase while the stocks exploited with low effort tend to decrease as compared to the case $\alpha=0$. This effect reduces the expected differences in harvest between agents employing high and low efforts with the high exploitation effort providing a slightly higher harvest. The probability for an agent with low effort to choose the high effort when comparing both harvests will thus almost, but not completely, be random $P(E^-\rightarrow E^+)\gtrapprox0.5$ and the system therefore shows a tendency to approach a state with all nodes employing the high effort ($n_-(t_f)\rightarrow 0$), but with increasing variance, i.e., increasing $\sigma(n_-(t_f))$

It has been checked that a variation of the rewiring probability $p$ in the stock network does not have a substantial influence on the system's behavior (not shown). In contrast, the average degree in the resource network $K_R$ alters the numerically estimated value of the critical coupling strength $\alpha_{crit}$, an effect that we study in detail in the next section.

\subsection{Optimized diffusive flow for the ecological subsystem leads to globally undesirable state}
\begin{figure*}[b]
\includegraphics[width=0.7\textwidth]{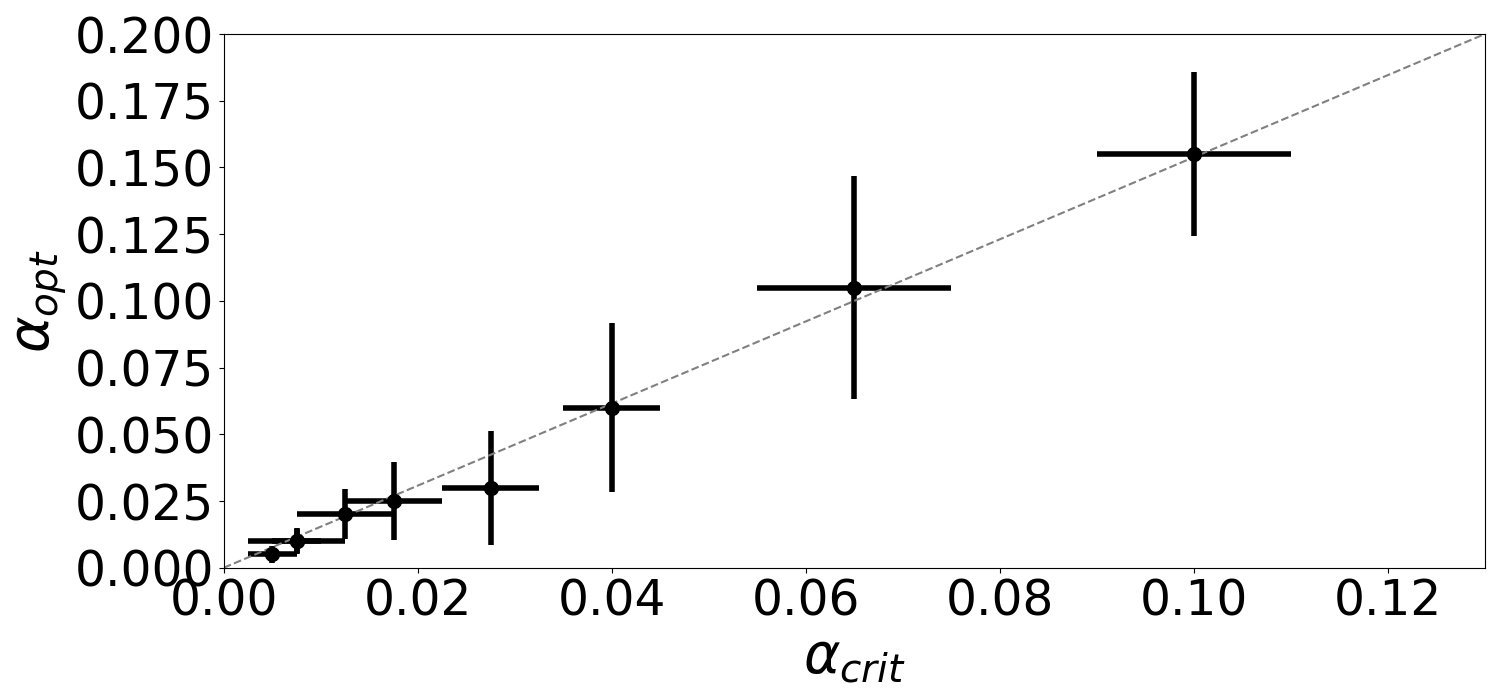}
\caption{\label{fig5}Optimal coupling coefficient $\alpha_{opt}$ as a function of the critical coupling coefficient $\alpha_{crit}$. The diagonal line indicates an approximate linear relationship.}
\end{figure*}
\label{sec:optimal_coupling_is_critical}

We now ultimately study how the critical coupling strength $\alpha_{crit}$ above which all agents choose a high effort for all considered social update times is related to the optimal coupling strength $\alpha_{opt}$ that we identified in Section \ref{resource_eq}. Particularly, we previously found that for $K_R=20$ the $\alpha_{crit}(K_R=20)\approx 0.0225$ (Fig.~\ref{fig3}a), while the optimal coupling strength has been estimated at $\alpha_{opt}(K_R=20)\approx 0.025$ (Fig. \ref{fig2}), thus appearing to be comparatively close to $\alpha_{crit}(K_R=20)$.

To further investigate the relationship between $\alpha_{opt}$ and $\alpha_{crit}$ we vary $\alpha$, social update time $\Delta T$ and stock network average degree $K_R$ and compute again the fraction of agents employing a low effort at consensus $\langle n_-(t_f)\rangle$ over ensembles of $n=50$ simulations per set of parameters, Fig.~\ref{fig4}. For each choice of the average degree $K_R$, a different $\alpha_{crit}$ is numerically identified (horizontal lines Fig.\ref{fig4}) and it appears that $\alpha_{crit}$ decreases with increasing stock network degree $K_R$, Fig.\ref{fig4}. We have additioanlly performed similar calculations for varying rewiring probabilities $p_R$ (not shown) and found that only the average degree has a significant influence on the critical coupling strength $\alpha_{crit}$. We have made similar observations with respect to the sensitivity of the results on $K_R$ and $p$ when estimating  $\alpha_{opt}$ (see above). 

In particular, we find that $\alpha_{crit}$ and $\alpha_{opt}$ display a linear dependency with $\alpha_{crit}\approx\alpha_{opt}$ for varying values of $K_R$,  Fig.~\ref{fig5}. As shown in Sec.~\ref{resource_eq} above, $\alpha_{opt}$ corresponds to a value at which the stocks exploited with high effort are best protected against depletion, thus optimizing the stocks' resilience against over-exploitation. However, this $\alpha_{opt}$ coincides with $\alpha_{crit}$ at or above which all agents in the social network layer choose a high effort. This leads to a critical state for the coupled socio-ecological system as a whole by causing all stocks to ultimately collapse and approach their undesired fixed point $s_0=0$, an effect that we denote here as the \textit{tragedy of the optimizer}.

This observed paradox or tragedy, i.e., a coupling strength that is optimal for the resource network leads to a likely collapse of the entire coupled socio-ecological system, has an intuitive explanation. Recall again that the optimal coupling coefficient $\alpha_{opt}$ was chosen to provide the largest possible protection against depletion of stocks which are exploited with high effort $E^+$ through a diffusive inflow of stocks which are exploited with low effort $E^-$. This implies an advantage for agents employing $E^+$, since their over-exploitation of stocks is compensated by the inflow of stocks from nodes harvested with $E^-$. Further increasing $\alpha>\alpha_{opt}$ amplifies this advantage and causes stocks that are exploited with $E^-$ and $E^+$ to become more similar. At the same time the harvests of nodes with $E^+$ are therefore expected to exceed those of nodes with $E^-$. Hence, for all values of $\alpha\geq \alpha_{opt}$ the system is expected to converge to a state where all agents hold $E^+$ once consensus is reached regardless the choice of the social update time $\Delta T$, Fig.~\ref{fig3}a and Fig~\ref{fig4}. This then also implies that this previously detected $\alpha_{opt}$ corresponds to the $\alpha_{crit}$ identified in the previous section.


\section{conclusion \& outlook}
\label{sec:conclusion}

Following up on previous studies of co-evolving socio-ecological networks with dynamic node states \cite{wiedermann2015macroscopic,barfuss_sustainable_2017, geier2019physics} we have introduced here a bi-layer network model that describes the interplay between a network of diffusively coupled resource stocks and a social network of agents updating their exploitation behaviors according to differences in harvest or payoff. 

We have first studied an implementation of the model with infinite social update time and from there it has been possible to define and explain the existence of an optimal coupling coefficient $\alpha_{opt}$. At this coupling coefficient the diffusive flow between the individual stocks in the network is optimized so that highly exploited stocks are best protected against complete depletion through the inflow from moderately exploited stocks.  

From there, we have studied the cases of finite social update times, meaning that interactions do not only take place along diffusive pathways in the resource network, but also along social ties in the corresponding network of interacting agents. We have observed that for lower coupling coefficients $\alpha$ the system undergoes a transition from a state where most agents choose a high exploitation effort to one where most agents prefer a low exploitation effort. The precise location of that transition varies with increasing social update times, a finding that is also in accordance with previous modelling results \cite{wiedermann2015macroscopic}. We have additionally estimated a critical coupling coefficient $\alpha_{crit}$ above which all agents prefer a high exploitation effort independently of the social update time. Ultimately we have found a second phase transition for comparatively large coupling coefficients at which the second moment of the system's equilibrium state shows a sudden increase. This transition is explained from the fact that agents choose either a high or a low effort at similar probabilities with only a slight preference for the high exploitation effort. 

The comparison between $\alpha_{opt}$ and $\alpha_{crit}$ has revealed that they are very similar across various choices of the average degree in the resource network, thus leading to what we denote a \textit{tragedy of the optimizer}. This tragedy implies that a coupling coefficient which appears to be optimal for the stock network on its own leads to a critical, hence undesired, transition in the coupled socio-ecological system as a whole. This effect shows that a measure which can stabilize the dynamics within a certain part of a larger co-evolutionary system may unexpectedly cause the emergence of new undesired globally stable states. This observation conceptually underlines the importance of a comprehensive approach for managing socio-ecological systems, since stabilizing effects that focus on a single sub-system may be detrimental for the system as a whole. 

The proposed model can be cautiously applied to a broader real-world context. In that sense, the here studied stock network could be considered to represent diffusively linked renewable natural resources, such as fish in a network of lakes and rivers. Agents with a high exploitation effort represent unsustainably acting individuals or communities, while agents with a low exploitation effort preserving natural resources long-term represent sustainably acting ones. Natural resources can be more resilient against over-exploitation when they are connected \cite{PERKINS2019,lundberg2003mobile}, and therefore protected, by healthy ecosystems around them, as is shown by the existence of an optimal coupling strength. The unsustainable harvesting of a natural resource at one location may then not be immediately experienced by other individuals that are managing this resource because surrounding ecological systems can to some degree compensate effects of over-exploitation through additional inflow. Unsustainable harvesting thus appears more attractive to shortsighted agents exploiting highly connected systems of ecological resources, which in its most extreme form of a common pool resource can lead to a {\em tragedy of the commons} \cite{hardin1968,ostrom}. As a consequence agents will tend to adopt harvesting strategies that (in the long-run) cannot be compensated by the increased resilience of the natural resource which ultimately increases the likelihood to overstretch planetary boundaries \cite{boundaries} through over-exploitation.

The described \textit{tragedy of the optimizer} raises challenging questions for future research. It should be investigated whether such a tragedy can also be observed in other types of coupled socio-ecological systems, such as systems with more than two network layers, including, e.g, a form of external management such as a governance layer~\cite{geier2019physics}. In addition the analysis should be broadened to also include other network topologies, such as scale-free or small-world networks or even real-world ecological networks, like river or lake ecosystems. Such an analysis could elucidate a different type of potential management strategies in such socio-ecological system, e.g., the creation of new connections between nodes in the resource networks, as is already done in other areas of study, such as power grid networks~\cite{schultz2014random}. Another approach for expanding the model would be to allow the harvesting agents to move from one resource to another, which has been shown to be a promising candidate to improve cooperation and, hence, potentially the convergence into a sustainable state~\cite{nag2020cooperation}.

Future work should also aim to cautiously compare the obtained model results to data from real-world phenomena where human intervention aiming to optimize the resilience of a natural resource failed and possibly led to adverse effects that could be linked to short-sighted optimization \cite{berkes2010shifting}. 

\begin{acknowledgments}
This work was developed in the context of the COPAN collaboration at the Potsdam Institute for Climate Impact Research (PIK).T.H. was supported by the German Research Foundation (DFG) (MU4430/2-1). M.W. is supported by the Leibniz Association (project DOMINOES).J.K. was supported by the Russian Ministry of Science and Education Agreement No. 075-15-2020-808. The authors gratefully acknowledge the European Regional Development Fund (ERDF), the German Federal Ministry of Education and Research and the Land Brandenburg for providing resources on the high-performance computer system at PIK.
\end{acknowledgments}

    


%

\end{document}